\documentclass[%
reprint,
superscriptaddress,
 amsmath,amssymb,
 aps,
 longbibliography,
floatfix,
]{revtex4-1}

\usepackage{graphicx}
\usepackage{dcolumn}
\usepackage{bm}
\usepackage{hyperref}
\usepackage{color}
\graphicspath{ {../Figures/} }
\begin{document}

\title{Electric field tunable superconductor-semiconductor coupling in Majorana nanowires}% Force line breaks with \\

\author{Michiel W. A. de Moor}
  \thanks{These authors contributed equally to this work.}%\\
\affiliation{%
 QuTech and Kavli Institute of NanoScience, Delft University of Technology, 2600 GA Delft, The Netherlands
}%
\author{Jouri D. S. Bommer}
  \thanks{These authors contributed equally to this work.}%\\
\affiliation{%
 QuTech and Kavli Institute of NanoScience, Delft University of Technology, 2600 GA Delft, The Netherlands
}%
\author{Di Xu}
  \thanks{These authors contributed equally to this work.}%\\
\affiliation{%
 QuTech and Kavli Institute of NanoScience, Delft University of Technology, 2600 GA Delft, The Netherlands
}%
\author{Georg W. Winkler}
\affiliation{%
 Station Q, Microsoft Research, Santa Barbara, California 93106-6105, USA
}%
\author{Andrey E. Antipov}
\affiliation{%
 Station Q, Microsoft Research, Santa Barbara, California 93106-6105, USA
}%
\author{Arno Bargerbos}
\affiliation{%
 QuTech and Kavli Institute of NanoScience, Delft University of Technology, 2600 GA Delft, The Netherlands
}%
\author{Guanzhong Wang}
\affiliation{%
 QuTech and Kavli Institute of NanoScience, Delft University of Technology, 2600 GA Delft, The Netherlands
}%
\author{Nick van Loo}
\affiliation{%
 QuTech and Kavli Institute of NanoScience, Delft University of Technology, 2600 GA Delft, The Netherlands
}%

\author{Roy L. M. Op het Veld}
\affiliation{%
 QuTech and Kavli Institute of NanoScience, Delft University of Technology, 2600 GA Delft, The Netherlands
}%
\affiliation{%
 Department of Applied Physics, Eindhoven University of Technology, 5600 MB Eindhoven, The Netherlands
}%
\author{Sasa Gazibegovic}
\affiliation{%
 QuTech and Kavli Institute of NanoScience, Delft University of Technology, 2600 GA Delft, The Netherlands
}%
\affiliation{%
 Department of Applied Physics, Eindhoven University of Technology, 5600 MB Eindhoven, The Netherlands
}%
\author{Diana Car}
\affiliation{%
 QuTech and Kavli Institute of NanoScience, Delft University of Technology, 2600 GA Delft, The Netherlands
}%
\affiliation{%
 Department of Applied Physics, Eindhoven University of Technology, 5600 MB Eindhoven, The Netherlands
}%
\author{John A. Logan}
\affiliation{%
 Materials Department, University of California, Santa Barbara, California 93106, USA
}%
\author{Mihir Pendharkar}
\affiliation{%
 Electrical and Computer Engineering, University of California, Santa Barbara, California 93106, USA
}%
\author{Joon Sue Lee}
\affiliation{%
 Electrical and Computer Engineering, University of California, Santa Barbara, California 93106, USA
}%
\author{Erik P. A. M. Bakkers}
\affiliation{%
 QuTech and Kavli Institute of NanoScience, Delft University of Technology, 2600 GA Delft, The Netherlands
}%
\affiliation{%
 Department of Applied Physics, Eindhoven University of Technology, 5600 MB Eindhoven, The Netherlands
}%
\author{Chris J. Palmstr\o m}
\affiliation{%
 Materials Department, University of California, Santa Barbara, California 93106, USA
}%
\affiliation{%
 Electrical and Computer Engineering, University of California, Santa Barbara, California 93106, USA
}%
\author{Roman M. Lutchyn}
\affiliation{%
 Station Q, Microsoft Research, Santa Barbara, California 93106-6105, USA
}%
\author{Leo P. Kouwenhoven}
\affiliation{%
 QuTech and Kavli Institute of NanoScience, Delft University of Technology, 2600 GA Delft, The Netherlands
}%
\affiliation{%
 Microsoft Station Q at Delft University of Technology, 2600 GA Delft, The Netherlands
}%
\author{Hao Zhang}
 \email{H.Zhang-3@tudelft.nl}
\affiliation{%
 QuTech and Kavli Institute of NanoScience, Delft University of Technology, 2600 GA Delft, The Netherlands
}
\date{\today}

\begin{abstract}
\indent We study the effect of external electric fields on superconductor-semiconductor coupling by measuring the electron transport in InSb semiconductor nanowires coupled to an epitaxially grown Al superconductor. We find that the gate voltage induced electric fields can greatly modify the coupling strength, which has consequences for the proximity induced superconducting gap, effective g-factor, and spin-orbit coupling, which all play a key role in understanding Majorana physics. We further show that level repulsion due to spin-orbit coupling in a finite size system can lead to seemingly stable zero bias conductance peaks, which mimic the behavior of Majorana zero modes. Our results improve the understanding of realistic Majorana nanowire systems.
\end{abstract} 

\maketitle
\section{Introduction}
\indent The hybrid superconductor-semiconductor nanowire system is the prime candidate to realize, control, and manipulate Majorana zero modes (MZMs) for topological quantum information processing~\cite{NayakRevModPhys2008,PluggeNPJ2017,KarzigPRB2017}. Majorana zero modes can be engineered in these hybrid nanowire systems by combining the one dimensional nature of the nanowire, strong spin-orbit coupling, superconductivity, and appropriate external electric (to control the chemical potential) and magnetic fields (to control the Zeeman energy) to drive the system into a topologically non-trivial phase~\cite{LutchynPRL2010,OregPRL2010}. To induce superconductivity in the semiconductor nanowire, it needs to be coupled to a superconductor. The electronic coupling between the two systems turns the nanowire superconducting~\cite{deGennesProximityEffect}, known as the proximity effect. Following this scheme, the first signatures of MZMs were observed in these hybrid systems, characterized by a zero bias peak (ZBP) in the tunneling conductance spectrum~\cite{MourikScience2012,DasNatPhys2012,DengNanoLett2012,ChurchillPRB2013}. Since then, significant progress has been made in Majorana experiments~\cite{GulBallisticMajorana2017,DengScience2016,ZhangNature2018,LutchynReview2018}, enabled by more uniform coupling between the superconductor and semiconductor nanowire. This has been achieved by improved interface engineering: through careful ex situ processing~\cite{GulHardGap2017,ZhangBallisticSC2017,GillarXiv2018}, by depositing the superconductor on the nanowires in situ~\cite{KrogstrupNatMat2015,ChangNatNano2015}, and a combination of in situ and ex situ techniques~\cite{SasaNature2017}, finally leading to the quantization of the Majorana conductance~\cite{ZhangNature2018}.\\
\indent However, the treatment of the superconductor-semiconductor coupling in the interpretation of experiments is often oversimplified. This coupling has recently been predicted to depend substantially on the confinement induced by external electric fields~\cite{AntipovarXiv2018}. In this work, we experimentally show that the superconductor-semiconductor coupling, as parameterized by the induced superconducting gap, is affected by gate induced electric fields. Due to the change in coupling, the renormalization of material parameters is altered, as evidenced by a change in the effective g-factor of the hybrid system. Furthermore, the electric field is shown to affect the spin-orbit interaction, revealed by a change in the level repulsion between Andreev states. Our experimental findings are corroborated by numerical simulations.
\section{Experimental set-up}
\begin{figure}[htbp]
		\includegraphics[width=8.6cm]{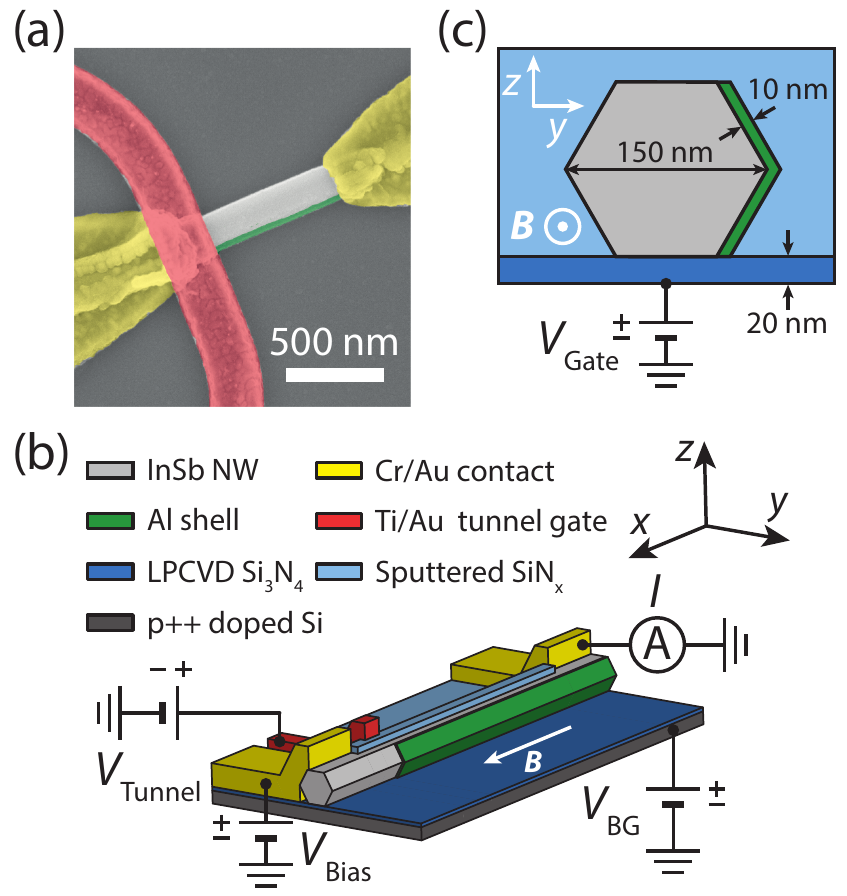}
		\centering
		\caption{\textbf{Device schematics}. (\textbf{a}) SEM of device A, with InSb nanowire in gray, superconducting aluminum shell in green, Cr/Au contacts in yellow, and local tunnel gate in red. Scale bar is 500 nm. (\textbf{b}) Schematic of experimental set-up. The substrate acts as a global back gate. The magnetic field is applied along the nanowire direction ($x$-axis). (\textbf{c}) Geometry used in the numerical simulations. A uniform potential $V_{\mathrm{Gate}}$ is applied as a boundary condition at the interface between substrate and dielectric. The superconductor (green) is kept at a fixed potential, which is set by the work function difference at the superconductor-semiconductor interface.}
\end{figure}
\begin{figure*}[htbp]
		\includegraphics[width=17.8cm]{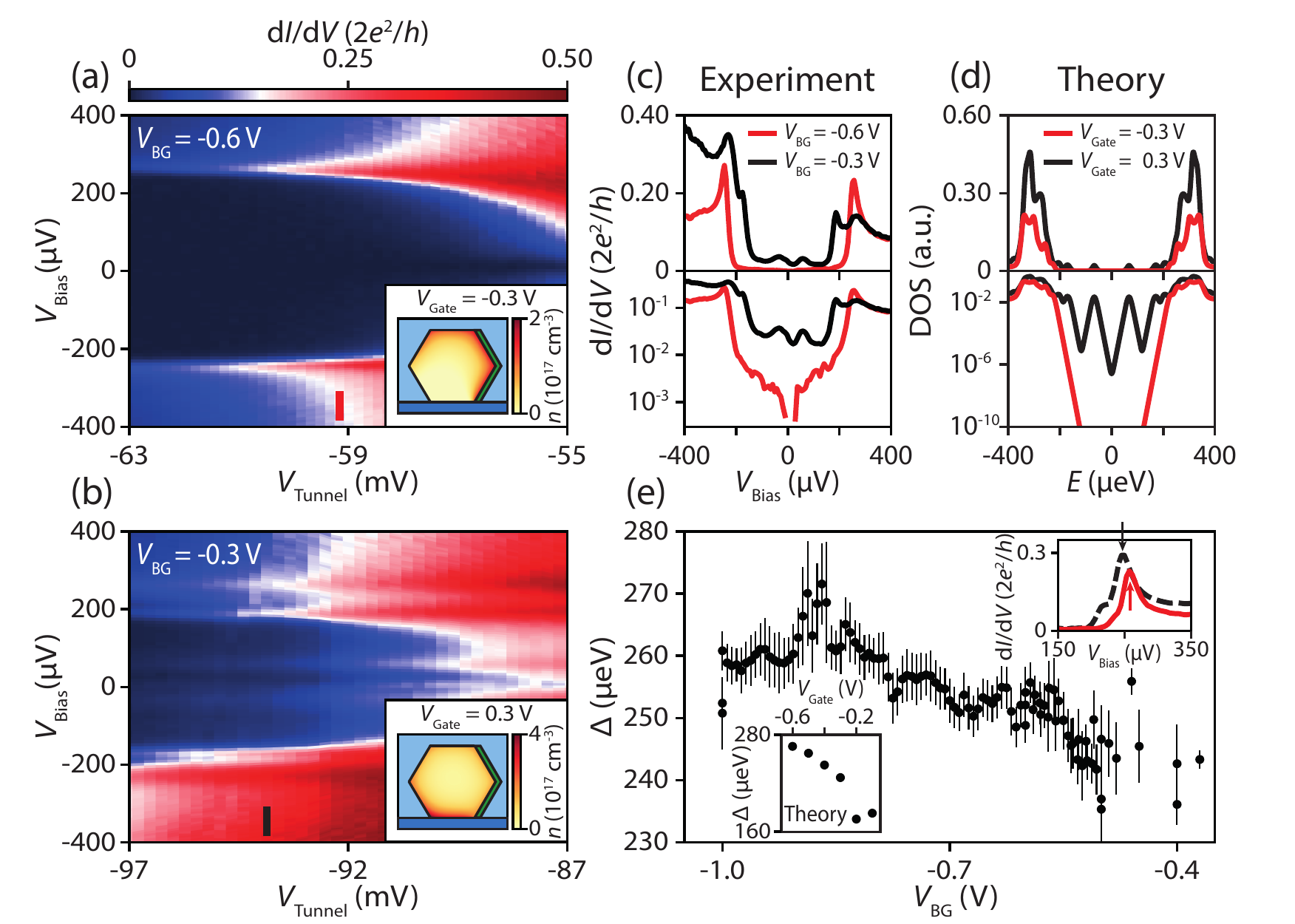}
		\centering
		\caption{\textbf{Gate dependence of the induced superconducting gap.} (\textbf{a},\textbf{b}) Differential conductance d$I$/d$V$ measured in device A as a function of $V_{\mathrm{Bias}}$ and $V_{\mathrm{Tunnel}}$ for $V_{\mathrm{BG}}$ = -0.6\,V (\textbf{a}) and $V_{\mathrm{BG}}$ = -0.3\,V (\textbf{b}). Insets show the calculated electron density in the wire for $V_{\mathrm{Gate}}$ = -0.3\,V and $V_{\mathrm{Gate}}$ = 0.3\,V, respectively. (\textbf{c}) Line-cuts from (a) and (b), indicated by the colored bars, in linear (top) and logarithmic (bottom) scale. (\textbf{d}) Calculated DOS for the density profiles shown in the insets of (a) and (b), shown in red and black, respectively. (\textbf{e}) Induced gap magnitude $\Delta$ as a function of $V_{\mathrm{BG}}$, showing a decrease for more positive gate voltages. Top right inset: line traces showing the coherence peak position (indicated by the arrow) for $V_{\mathrm{BG}}$ = -0.6\,V (solid red line) and $V_{\mathrm{BG}}$ = -0.4\,V (dashed black line). Bottom left inset: induced gap from the calculated DOS as a function of $V_{\mathrm{Gate}}$, consistent with the experimental observation.}
\end{figure*}
\indent We have performed tunneling spectroscopy experiments on four InSb-Al hybrid nanowire devices, labeled A-D, all showing consistent behaviour. The nanowire growth procedure is described in ref.~\cite{SasaNature2017}. A scanning electron micrograph (SEM) of device A is shown in Fig.~1(a). Figure~1(b) shows a schematic of this device and the measurement set-up. For clarity, the wrap-around tunnel gate, tunnel gate dielectric and contacts have been removed on one side. A normal-superconductor (NS) junction is formed between the part of the nanowire covered by a thin shell of aluminum (10 nm thick, indicated in green, S), and the Cr/Au contact (yellow, N). The transmission of the junction is controlled by applying a voltage $V_{\mathrm{Tunnel}}$ to the tunnel gate (red), galvanically isolated from the nanowire by 35 nm of sputtered SiN$_x$ dielectric. The electric field is induced by a global back gate voltage $V_{\mathrm{BG}}$, except in the case of device B, where this role is played by the side gate voltage $V_{\mathrm{SG}}$ (not shown in Fig. 1)~\cite{Supplement}. To obtain information about the density of states in the proximitized nanowire, we measure the differential conductance d$I$/d$V_{\mathrm{Bias}}$ as a function of applied bias voltage $V_{\mathrm{Bias}}$. In the following, we will label this quantity as d$I$/d$V$ for brevity. A magnetic field is applied along the nanowire direction ($x$-axis in Figs.~1(b),1(c)). All measurements are performed in a dilution refrigerator with a base temperature of 20\,mK.
\section{Theoretical model}
\indent The device geometry used in the simulation is shown in Fig.~1(c). We consider a nanowire oriented along the $x$-direction, with a hexagonal cross-section in the $yz$-plane. The hybrid superconductor-nanowire system is described by the Bogoliubov-de Gennes Hamiltonian
\begin{equation}
  \begin{aligned}
    H=&\left[\frac{\hbar^2 \mathbf k^2}{2m^*}-\mu-e\phi\right]\tau_z%+\bm\alpha
                                %\cdot \mathbf k \times \bm\sigma
                                %\tau_z\\
    +\alpha_y (k_z \sigma_x - k_x \sigma_z)\tau_z\\
    &+\alpha_z (k_x\sigma_y - k_y \sigma_x)\tau_z
        +\frac{1}{2}g\mu_\mathrm{B} B\sigma_x+\Delta \tau_x.
    \label{eq:ham}
  \end{aligned}
\end{equation}
The first term contains contributions from the kinetic energy and the chemical potential, as well as the electrostatic potential $\phi$. The second and third terms describe the Rashba spin-orbit coupling, with the coupling strength $\alpha_y$ ($\alpha_z$) depending on the $y$-component ($z$-component) of the electric field. The Zeeman energy contribution, proportional to $g$, the Land\'{e} g-factor, is given by the fourth term. Finally, the superconducting pairing $\Delta$ is included as the fifth term. All material parameters are position dependent, taking different values in the InSb nanowire and the Al superconductor~\cite{Supplement}.\\
\indent If the coupling between the superconductor and semiconductor is small (compared to the bulk gap of the superconductor $\Delta$, known as weak coupling), superconductivity can be treated as a constant pairing potential term in the nanowire Hamiltonian, with the induced superconducting gap being proportional to the coupling strength~\cite{VolkovPhysicaC1995}. However, if the coupling becomes strong, the wave functions of the two materials hybridize, and the superconductor and semiconductor have to be considered on equal footing~\cite{ReegPRB2018}. We achieve this by solving the Schr\"{o}dinger equation in both materials simultaneously. When desired, the orbital effect of the magnetic field is added via Peierls substitution~\cite{NijholtPRB2016}. The simulations are performed using the \texttt{kwant} package~\cite{kwant}.\\
\indent The electrostatic potential in the nanowire cross-section is calculated from the Poisson equation, assuming an infinitely long wire. We use a fixed potential $V_{\mathrm{Gate}}$ as a boundary condition at the dielectric-substrate interface. The superconductor enters as the second boundary condition, with a fixed potential to account for the work function difference between superconductor and semiconductor~\cite{VuikNPJ2016}. We approximate the mobile charges in the nanowire by a 3D electron gas (Thomas-Fermi approximation). It has been demonstrated that the potentials calculated using this approximation give good agreement with results obtained by self-consistent Schr\"{o}dinger-Poisson simulations~\cite{MikkelsenarXiv2018}. The calculated potential for a given $V_{\mathrm{Gate}}$ is then inserted into the Hamiltonian~(\ref{eq:ham}).\\
\indent By solving the Schr\"{o}dinger equation for a given electrostatic environment, we can see how the gate potential alters the electronic states in the nanowire, how they are coupled to the superconductor, and how this coupling affects parameters such as the induced gap, effective g-factor, and spin-orbit energy.
\begin{figure*}[htbp]
		\includegraphics[width=17.8cm]{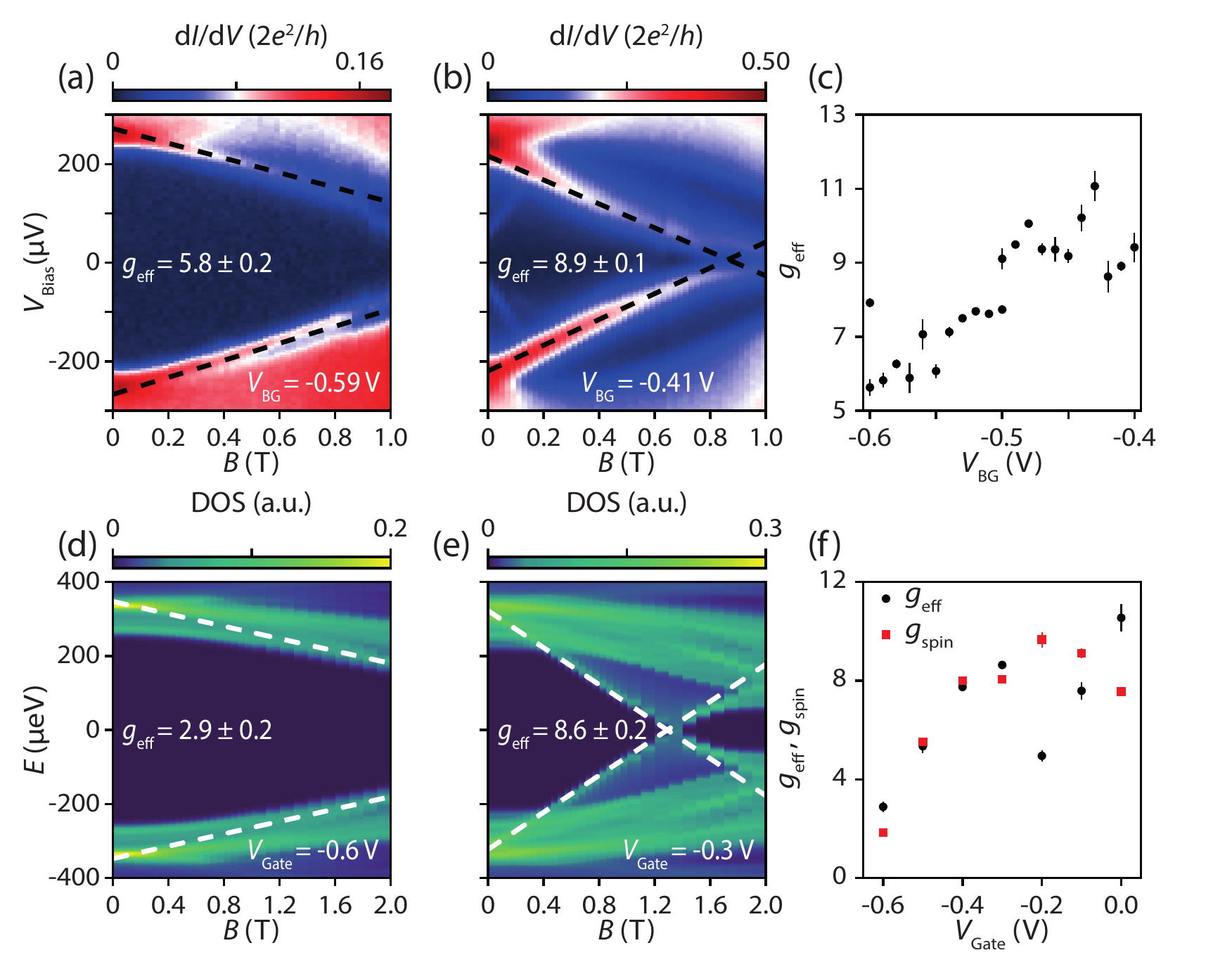}
		\centering
		\caption{\textbf{Effective g-factor.} (\textbf{a},\textbf{b}) d$I$/d$V$ measured in device A as a function of applied bias voltage $V_{\mathrm{Bias}}$ and magnetic field $B$ for $V_{\mathrm{BG}}$ = -0.59\,V and $V_{\mathrm{BG}}$ = -0.41\,V, respectively. The effective g-factor is extracted from a linear fit of the lowest energy state dispersion (dashed lines). (\textbf{c}) $g_{\mathrm{eff}}$ as a function of $V_{\mathrm{BG}}$, showing an increase as the gate voltage becomes more positive. Data from device A. (\textbf{d},\textbf{e}) Simulated DOS in the nanowire as a function of magnetic field for $V_{\mathrm{Gate}}$ = -0.6\,V and $V_{\mathrm{Gate}}$ = -0.3\,V, respectively. (\textbf{f}) Extracted $g_{\mathrm{eff}}$ (based on lowest energy state in the spectrum, black circles) and $g_{\mathrm{spin}}$ (based on the spectrum at $k$ = 0, red squares) from the simulation.}
\end{figure*}
\section{Gate voltage dependence of the induced superconducting gap}
When the transmission of the NS-junction is sufficiently low (i.e., in the tunneling regime), the differential conductance d$I$/d$V$ is a direct measure of the density of states (DOS) in the proximitized nanowire~\cite{BardeenPRL1961}. In Fig.~2(a), we plot d$I$/d$V$ measured in device A as a function of applied bias voltage $V_{\mathrm{Bias}}$ and tunnel gate voltage $V_{\mathrm{Tunnel}}$, for $V_{\mathrm{BG}}$ = -0.6\,V. In the low transmission regime, we resolve the superconducting gap $\Delta$ around 250 $\mathrm{\mu}$eV, indicated by the position of the coherence peaks. The ratio of sub-gap to above-gap conductance (proportional to the normal state transmission of the junction, $T$) follows the behavior expected from BTK theory~\cite{BTK,BeenakkerPRB1992}, indicating the sub-gap conductance is dominated by Andreev reflection processes (proportional to $T^2$). This is generally referred to as a hard gap. However, for more positive back gate voltages, the sub-gap conductance is larger and shows more resonances, as is illustrated in Fig.~2(b) for $V_{\mathrm{BG}}$ = -0.3\,V. Fig.~2(c) shows line traces taken at a similar transmission (above-gap conductance) for both cases. The sub-gap conductance for $V_{\mathrm{BG}}$ = \nobreakdash-0.3\,V (black line) exceeds that of the hard gap case (red line) by an order of magnitude. This is indicative of a surplus of quasi-particle states inside the gap, referred to as a soft gap.\\
\indent The gate voltage induced transition from soft to hard gap is generically observed in multiple devices. To understand this phenomenology, we calculate the electron density in the nanowire cross-section for different values of $V_{\mathrm{Gate}}$. Because the charge neutrality point in our devices is unknown,  there is a difference between the gate voltages used in the experiment and the values of $V_{\mathrm{Gate}}$ used in the simulation. By comparing the transition point between hard and soft gaps in the experiment and the simulation, we estimate that the experimental gate voltage range 
-0.6\,V~\textless~$V_{\mathrm{BG}}$~\textless~-0.4\,V roughly corresponds to the simulated gate voltage range -0.4\,V~\textless~$V_{\mathrm{Gate}}$~\textless~-0.2\,V.\\
\indent For more negative $V_{\mathrm{Gate}}$, the electric field from the gate pushes the electrons towards interface with the superconductor (inset of Fig.~2(a)). We solve the Schr\"{o}dinger equation for the calculated electrostatic potential and find that this stronger confinement near the interface leads to a stronger coupling. This results in a hard gap, as illustrated by the calculated energy spectrum (Fig.~2(d), red line). However, for more positive voltages, the electrons are attracted to the back gate, creating a high density pocket far away from the superconductor (inset of Fig.~2(b)). These states are weakly coupled to the superconductor, as demonstrated by a soft gap structure (Fig.~2(d), black line). We can therefore conclude that the electron tunneling between the semiconductor and the superconductor is strongly affected by the gate potential.\\
\indent The change in superconductor-semiconductor coupling does not just affect the hardness, but also the size of the gap. For each back gate voltage, we fit the BCS-Dynes expression~\cite{Dynes} for the DOS in order to extract the position of the coherence peaks, giving the gap size $\Delta$. The results are shown in Fig.~2(e). As $V_{\mathrm{BG}}$ becomes more positive, the superconductor-semiconductor coupling becomes weaker, reducing the size of the gap. From $V_{\mathrm{BG}}$ \textgreater \,-0.4\,V onward it becomes difficult to accurately determine the gap, as it tends to become too soft and the coherence peaks are not always clearly distinguishable. The top right inset shows the shift of the coherence peak (indicated by the arrows) to lower bias voltage as $V_{\mathrm{BG}}$ is increased. The lower left inset shows the extracted coherence peak position from the numerical simulations, showing the same trend.
\section{Effective g-factor}
\indent As the electric field induced by the back gate clearly has an important effect on the hybridization between the nanowire and the superconductor, we now look at the effect this has on the Zeeman term in the Hamiltonian. This term affects the energy dispersion of spinful states in a magnetic field. We study the dispersion of the states in the nanowire by measuring d$I$/d$V$ in device A as a function of applied bias voltage and magnetic field, as shown in Fig.~3(a) and Fig.~3(b). We define the effective g-factor as $g_{\mathrm{eff}} = \frac{2}{\mu_B} \frac{\Delta E}{\Delta B}$, with $\frac{\Delta E}{\Delta B}$ the average slope of the observed peak in the differential conductance as it disperses in magnetic field. This effective g-factor is different from the pure spin g-factor $g_{\mathrm{spin}}$, as the dispersion used to estimate $g_{\mathrm{eff}}$ is generally not purely linear in magnetic field, and has additional contributions from the spin-orbit coupling, magnetic field induced changes in chemical potential, and orbital effects~\cite{VuikNPJ2016,AntipovarXiv2018,Winkler2017}. The effective g-factor is the parameter which determines the critical magnetic field required to drive the system through the topological phase transition~\cite{DasSarmaPRB2011}. We obtain the slope $\frac{\Delta E}{\Delta B}$ from a linear fit (shown as black dashed lines in Fig.~3(a),(b)) of the observed peak position. Fig.~3(c) shows the extracted $g_{\mathrm{eff}}$ for device A, with more positive back gate voltages leading to larger $g_{\mathrm{eff}}$ (visible as a steeper slope). A similar result has recently been reported in hybrid InAs-Al nanowires~\cite{VaitiekenasarXiv2017}.\\
\indent We use our numerical model to calculate the DOS in the nanowire as a function of applied magnetic field, shown in Fig.~3(d) and Fig.~3(e). From the calculated spectrum, we apply the same procedure used to fit the experimental data to extract $g_{\mathrm{eff}}$ (white dashed lines). The results for different values of $V_{\mathrm{Gate}}$ are given in Fig.~3(f) as black circles. The applied back gate voltage changes the hybridization of the states in the InSb ($\lvert g_{\mathrm{spin}}\rvert$ = 40~\cite{KammhuberNanoLett2016}) and the Al ($\lvert g_{\mathrm{spin}}\rvert$ = 2). As a more positive gate voltage increases the weight of the wave function in the InSb, we expect the renormalized g-factor to increase as the gate voltage is increased, consistent with the results of Fig.~3(c) and Fig.~3(f). \\
\indent To see how well $g_{\mathrm{eff}}$ describes the Zeeman term in the Hamiltonian, we turn our attention to the energy spectrum at $k$ = 0. At this point, the effect of spin-orbit coupling vanishes. If orbital effects are excluded, we can then define the pure spin g-factor as $g_{\mathrm{spin}} = \frac{2}{\mu_B} \frac{\Delta E(k=0)}{\Delta B}$. The resulting values for $g_{\mathrm{spin}}$ are shown as red squares in Fig.~3(f). By comparing the results for $g_{\mathrm{eff}}$ and $g_{\mathrm{spin}}$, we can conclude that when the lowest energy state has a momentum near $k$ = 0 (as is the case for $V_{\mathrm{Gate}} \textless$ -0.2\,V), the effect of spin-orbit coupling is negligible, and $g_{\mathrm{eff}}$ is a good proxy for the pure spin g-factor. However, when this is no longer the case, significant deviations can be observed, as is the case for $V_{\mathrm{Gate}} \geq$ -0.2\,V. As we expect the experimental gate voltage range of Fig.~3(c) to be comparable to values of $V_{\mathrm{Gate}}$\,\textless\,-0.2\,V, we conclude that the experimentally obtained $g_{\mathrm{eff}}$ is a reasonable approximation of $g_{\mathrm{spin}}$ in this parameter regime. However, we stress once more that in general, one needs to be careful when interpreting the $g_{\mathrm{eff}}$ extracted from experimental data as the g-factor entering the Hamiltonian in the Zeeman term.\\
\indent The increasing trend of $g_{\mathrm{eff}}$ does not change when the orbital effect of magnetic field is considered~\cite{Supplement}. However, there is a significant increase in the predicted values, in agreement with previous findings for InAs nanowires~\cite{Winkler2017}. These values are larger than the ones generally observed in our experiment, suggesting that the orbital effect is not a dominant mechanism in determining the effective g-factor in these devices. We note that the data from this device was taken solely in the hard gap regime, where one expects a strong confinement near the superconductor. This suppresses the orbital contribution of the magnetic field. Another possible explanation for the discrepancy between the results of the simulation and the experimental data is an overestimation of the density in the nanowire, as higher sub-bands have a stronger contribution from the orbital effect. Minimizing the orbital effect is desirable for Majorana physics, as the orbital contributions of the magnetic field are detrimental to the topological gap~\cite{NijholtPRB2016}.
\section{Level repulsion due to spin-orbit coupling}
\begin{figure*}[htbp]
		\includegraphics[width=17.8cm]{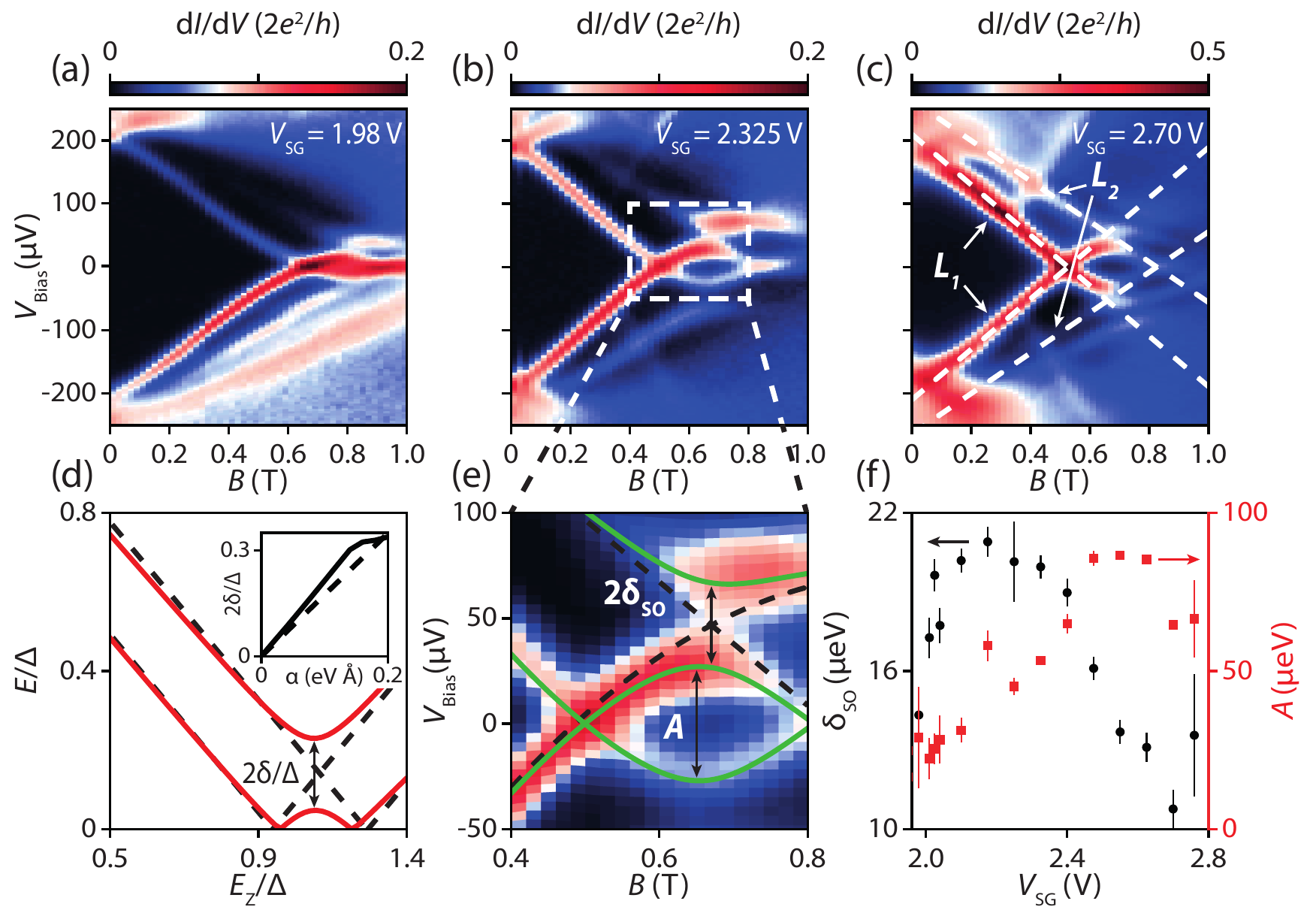}
		\centering
		\caption{\textbf{Spin-orbit coupling induced level repulsion.} (\textbf{a}-\textbf{c}) d$I$/d$V$ as a function of $V_{\mathrm{Bias}}$ for device B, showing the dispersion of subgap states in magnetic field, for $V_{\mathrm{SG}}$ = 1.98\,V, 2.325\,V, and 2.70\,V, respectively. The two lowest energy states $L_1$, $L_2$, and their particle-hole symmetric partners are indicated by the white dashed lines. (\textbf{d}) Calculated low energy spectrum of the finite nanowire system as a function of the Zeeman energy $E_\mathrm{Z}$ for $\alpha$ = 0\,eV\,\AA (dashed black lines) and $\alpha$ = 0.1\,eV\,\AA (solid red lines), showing the opening of an energy gap 2$\delta$ due to spin-orbit coupling. Inset: the energy gap 2$\delta$ as a function of the Rashba $\alpha$ parameter (solid line), and the estimate 2$\delta$ = $\alpha\pi/l$ (dashed line), with $l$ the nanowire length. All energy scales are in units of the superconducting gap $\Delta$. (\textbf{e}) Zoom-in of the anti-crossing in (\textbf{b}), showing the splitting $A$ and the coupling strength $\delta_{\mathrm{SO}}$. Green solid lines indicate a fit of the anti-crossing, with the dashed black lines showing the uncoupled energy levels. (\textbf{f}) Coupling $\delta_{\mathrm{SO}}$ (black circles) and splitting $A$ (red squares) as a function of $V_{\mathrm{SG}}$, showing opposite trends for these parameters.}
\end{figure*}
\begin{figure*}[htbp]
		\includegraphics[width=17.8cm]{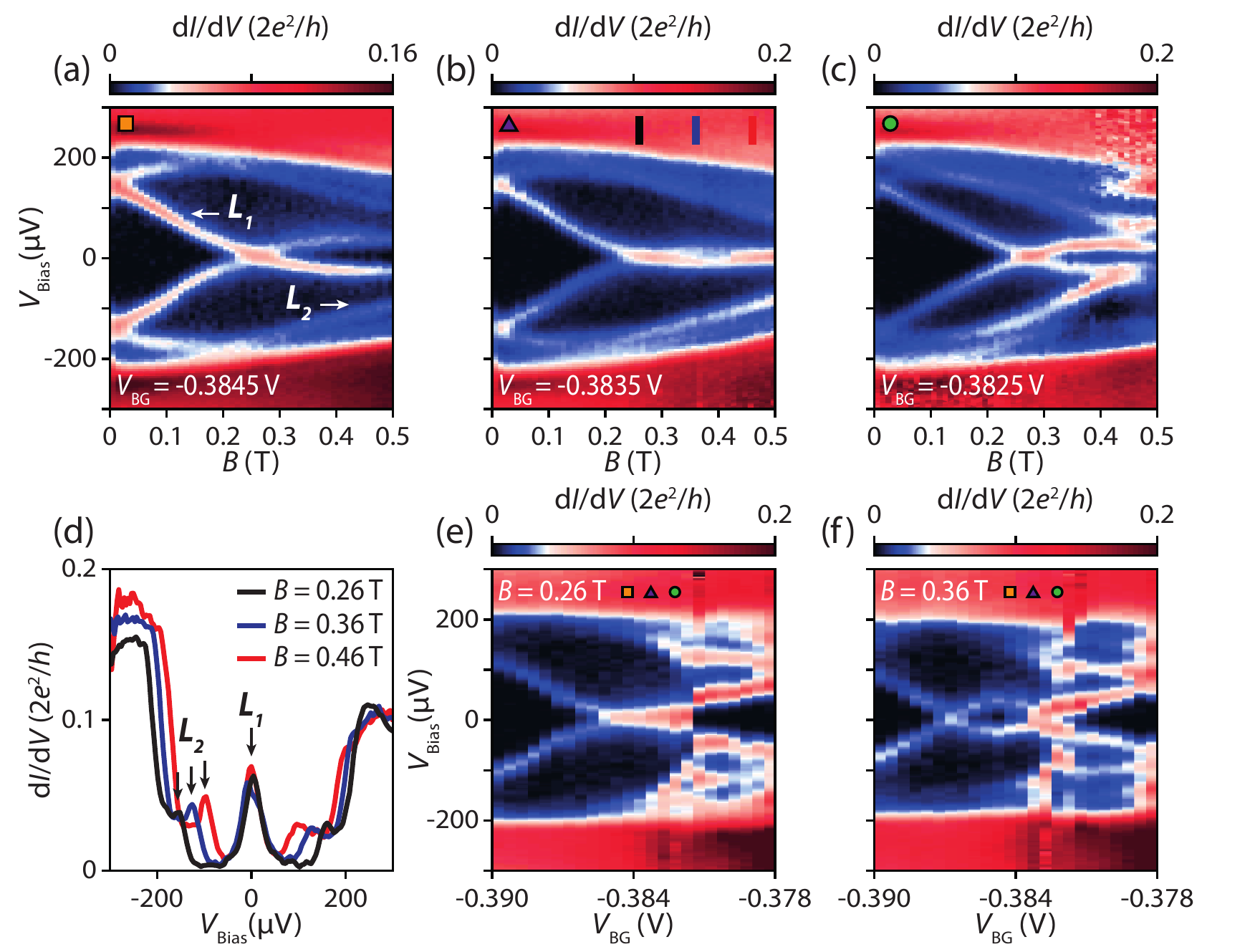}
		\centering
		\caption{\textbf{Zero bias pinning due to strong level repulsion.} (\textbf{a}-\textbf{c}) d$I$/d$V$ as a function of $V_{\mathrm{Bias}}$ for device A, showing the dispersion of $L_1$ and $L_2$ as a function of magnetic field for $V_{\mathrm{BG}}$ = -0.3845\,V, -0.3835\,V, and -0.3825\,V, respectively. (\textbf{d}) Line traces at magnetic fields indicated by the colored bars in (b), showing the stable pinning of $L_1$ to zero bias voltage. (\textbf{e},\textbf{f}) d$I$/d$V$ measured as a function of $V_{\mathrm{BG}}$ at fixed magnetic field $B$ = 0.26\,T and 0.36\,T, respectively. Gate voltages from (a), (b), and (c) are indicated by orange square, purple triangle, and green circle, respectively.}
\end{figure*}
\indent The last term in the Hamiltonian that remains to be explored describes the Rashba spin-orbit coupling. The strength of the spin-orbit coupling is determined by the parameter $\alpha$, which depends on the material (and thus, on the superconductor-semiconductor coupling), and the electric field~\cite{NittaPRL1997,vanWeperenPRB2015,ScherublInAsSOI2016}. Therefore, we expect that this term will be affected by the gate potential as well. In finite systems, the spin-orbit interaction can couple states with different orbitals and spins~\cite{StanescuPRB2013}. These states are thus no longer orthogonal to each other, and the spin-orbit mediated overlap between them causes energy splitting, leading to level repulsion~\cite{LeeNatureNano2014,vanHeckPRB2017,OFarrellarXiv2018}. This level repulsion, which is generic in class D systems in the presence of superconductivity, magnetic field and spin-orbit coupling~\cite{PikulinNPJ2012}, can be extracted from the low energy nanowire spectrum as measured by tunneling spectroscopy.\\
\indent In Figs.~4(a)-(c), we show the evolution of the level repulsion between the two lowest energy sub-gap states (labeled $L_1$ and $L_2$, as indicated by the white dashed lines in panel c) in device B. For these measurements, the global back gate is grounded, with the electric field being induced by applying a voltage to the side gate~\cite{Supplement}.\\
\indent We parameterize the level repulsion by two quantities: the coupling strength $\delta_{\mathrm{SO}}$, and the splitting $A$, defined as the maximum deviation of $L_1$ from zero energy after the first zero crossing. This splitting has previously been linked to the overlap between two MZM in a finite system~\cite{AlbrechtNature2016}. In Fig.~4(e), we zoom in on the anti-crossing feature in panel Fig.~4(b), showing the minimum energy difference between $L_1$ and $L_2$ (given by 2$\delta_{\mathrm{SO}}$) and the splitting $A$. We extract these parameters by a fit of the anti-crossing (solid green lines, with the uncoupled states shown by the dashed black lines)~\cite{Supplement}.\\
\indent Because we expect finite size effects to be relevant, we cannot use our previous theoretical model, as it is based on an infinitely long nanowire. Therefore, we modify the model to take into account the finite size of the nanowire system, and calculate the low energy spectrum for different values of the Rashba spin-orbit strength~\cite{Supplement}. In Fig.~4(d), we plot the two lowest energy states in the nanowire as a function of the Zeeman energy ($E_\mathrm{Z} = \frac{1}{2}g\mu_B B$), in units of the superconducting gap $\Delta$. If $\alpha$ = 0 (no spin-orbit coupling, dashed black lines), there is no coupling between the states, and no level repulsion occurs. However, if spin-orbit coupling is included (e.g., $\alpha$ = 0.1\,eV\,\AA, solid red lines), the levels repel each other, with the magnitude of the anti-crossing given by 2$\delta$. The level repulsion strength scales with $\alpha$ (inset of Fig.~4(d)), providing a way to estimate $\alpha$ based on the low energy spectrum using 2$\delta \sim \alpha\pi/l$, where $l$ is the length of the nanowire.\\
\indent In Fig.~4(f), we plot $\delta_{\mathrm{SO}}$ (black circles) and $A$ (red squares) as a function of the applied side gate voltage. The two parameters follow opposite trends, with $A$ being maximal when $\delta_{\mathrm{SO}}$ is minimal. When $\delta_{\mathrm{SO}}$ is larger, the levels repel each other more, leading to $L_1$ being pushed closer to zero energy, reducing the splitting $A$. When $V_{\mathrm{SG}}$\,\textless \,2.0\,V, both parameters become smaller with decreasing $V_{\mathrm{SG}}$. At this point, other states at higher energies become relevant for the lowest energy dispersion (a situation demonstrated in Fig.~4(a)), and our method to extract these parameters breaks down. We expect this method to be reliable when the energetically lowest two states can be clearly separated from the rest.\\ \indent Because $\delta_{\mathrm{SO}}$ depends not only on $\alpha$, but also on the details of the confinement potential, as well as the coupling to the superconductor, a precise estimate goes beyond the current approximations in our model. That being said, based on the observed magnitude of $\delta_{\mathrm{SO}}$ and our simulations of the finite nanowire system, we can estimate the Rashba parameter $\alpha$ to be around 0.1\,eV\,\AA~in this gate voltage range. This value is comparable to the values reported in InSb nanowire based quantum dots~\cite{NadjPergePRL2012}, and smaller than the values measured in weak anti-localization experiments~\cite{vanWeperenPRB2015}. A large value of $\alpha$ is beneficial for Majorana physics, as it determines the maximum size of the topological gap~\cite{SauPRB2012}.
\section{Zero Bias Peak in extended magnetic field range}
\indent In the previous sections, we have described the effect of the gate induced electric field on the various terms in the Hamiltonian (\ref{eq:ham}). As this Hamiltonian is known to describe Majorana physics, we now turn our attention to possible signatures of MZMs in this system. In particular, when 2$\delta_{\mathrm{SO}}$ becomes comparable to the energy of $L_2$, we find that $L_1$ can become pinned close to zero bias over an extended range in magnetic field, as demonstrated in Fig.~5(b) (data from device A). Fig.~5(d) shows that the state stays pinned to zero energy over a range of over 0.2\,T, corresponding to a Zeeman energy of over 300\,$\mathrm{\mu}$eV, which is larger than the induced gap. The stability of the ZBP in terms of the ratio of Zeeman energy to induced gap is comparable to the most stable ZBPs reported in literature~\cite{DengScience2016,GulBallisticMajorana2017}. When we fix the magnetic field to $B$ = 0.26\,T and change the back gate voltage (Fig.~5(e)), it appears that there is a stable ZBP over a few mV as well.\\
\indent We might be tempted to conclude that this stability implies this is a Majorana zero mode. However, if we change either the gate voltage (Fig.~5(a), Fig.~5(c)) or the magnetic field (Fig.~5(f)) a little bit, we observe that this stability applies only to very particular combinations of gate voltage and magnetic field. One should keep in mind that in a finite system, MZMs are not expected to be stable with respect to local perturbations if the system size is comparable to the Majorana coherence length, which is likely the case in our devices. This further complicates the determination of the origin of the observed peaks. As we find no extended region of stability, we conclude that it is unlikely that this state pinned to zero energy is caused by a topological phase transition. Rather, this seems to be due to a fine-tuned coincidence in which the repulsion between two states combined with particle-hole symmetry leads to one of the states being pinned to $E$ = 0. We reiterate that simply having a stable zero energy state over an extended range in magnetic field is not sufficient to make claims about robust Majorana modes~\cite{KellsSmoothPotential2012,PradaNS2012,LiuTrivialABSMajorana2017}. Further experimental checks, such as stability of the ZBP in an extended region of the parameter space spanned by the relevant gate voltages~\cite{GulBallisticMajorana2017}, as well as magnetic field, are required in order to assign a possible Majorana origin. 
\section{Conclusion \& Outlook}
\indent We have used InSb nanowires with epitaxial Al superconductor to investigate the effect of the gate voltage induced electric field on the superconductor-semiconductor coupling. This coupling is determined by the distribution of the wave function over the superconductor and semiconductor, and controls essential parameters of the Majorana Hamiltonian: the proximity induced superconducting gap, the effective g-factor, and spin-orbit coupling. Our observations show that the induced superconductivity, as parameterized by the hardness and size of the induced gap, is stronger when the electrons are confined to a region close to the superconductor. The stronger coupling leads to a lower effective g-factor. We also determine that the gate voltage dependence of the effective g-factor is dominated by the change in coupling to the superconductor, rather than by orbital effects of the magnetic field. Finally, we study the effect of level repulsion due to spin-orbit coupling. Appropriate tuning of the repulsion leads to level pinning to zero energy over extended parameter ranges, mimicking the behavior expected from MZMs. Our result deepens the understanding of a more realistic Majorana nanowire system. More importantly, it is relevant for the design and optimization of future advanced nanowire systems for topological quantum information applications.\\
\begin{acknowledgments}
\indent We thank J.G. Kroll, A. Proutski, and S. Goswami for useful discussions. This work has been supported by the European Research Council, the Dutch Organization for Scientific Research, the Office of Naval Research, the Laboratory for Physical Sciences, and Microsoft Corporation Station Q.
\end{acknowledgments}
\section*{Author contributions}
\indent M.W.A.d.M., J.D.S.B., D.X., and H.Z. fabricated the devices, performed the measurements, and analyzed the data. G.W.W., A.B., A.E.A., and R.M.L. performed the numerical simulations. N.v.L. and G.W. contributed to the device fabrication. R.L.M.o.h.V., S.G., and D.C. grew the InSb nanowires under the supervision of E.P.A.M.B.. J.A.L., M.P., and J.S.L. deposited the aluminum shell on the nanowires under the supervision of C.J.P.. L.P.K. and H.Z. supervised the project. M.W.A.d.M. and H.Z. wrote the manuscript with comments from all authors. M.W.A.d.M., J.D.S.B., and D.X. contributed equally to this work. Correspondence to H.Z. (H.Zhang-3@tudelft.nl).
%\bibliography{Electric_field_tuning_refs}

\end{document}